\documentclass[USenglish,oneside,twocolumn]{article}

\usepackage{algorithmic}
\usepackage{algorithm}
\usepackage[utf8]{inputenc}
\usepackage[big]{dgruyter_NEW}
\usepackage{multirow}
\usepackage{multicol}
\usepackage{tabularx}
\usepackage{graphicx}
\usepackage{url}
\usepackage{array}
\usepackage{amsmath}
\usepackage{booktabs}

\mathchardef\mhyphen="2D
\makeatletter
\def\plist@algorithm{Alg.\space}
\makeatother
 
\cclogo{\includegraphics{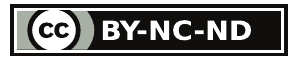}}
  
\begin{document}

  \author*[1]{James K Holland}

  \author[2]{Nicholas Hopper}



  \affil[1]{University of Minnesota, E-mail: holla556@umn.edu}

  \affil[2]{University of Minnesota, E-mail: hoppernj@umn.edu}



  \title{\huge RegulaTor: A Straightforward Website Fingerprinting Defense}

  \runningtitle{A Straightforward Website Fingerprinting Defense}


  \begin{abstract}
    {Website Fingerprinting (WF) attacks are used by local passive attackers to determine the destination of encrypted internet traffic by comparing the sequences of packets sent to and received by the user to a previously recorded data set. As a result, WF attacks are of particular concern to privacy-enhancing technologies such as Tor. In response, a variety of WF defenses have been developed, though they tend to incur high bandwidth and latency overhead or require additional infrastructure, thus making them difficult to implement in practice. Some lighter-weight defenses have been presented as well; still, they attain only moderate effectiveness against recently published WF attacks. In this paper, we aim to present a realistic and novel defense, RegulaTor, which takes advantage of common patterns in web browsing traffic to reduce both defense overhead and the accuracy of current WF attacks. In the closed-world setting, RegulaTor reduces the accuracy of the state-of-the-art attack, Tik-Tok, against comparable defenses from 66\% to 25.4\%. To achieve this performance, it requires limited added latency and a bandwidth overhead 39.3\% less than the leading moderate-overhead defense. In the open-world setting, RegulaTor limits a precision-tuned Tik-Tok attack to an $F_1$-score of .135, compared to .625 for the best comparable defense. }
\end{abstract}

  \journalname{Proceedings on Privacy Enhancing Technologies}
\DOI{Editor to enter DOI}
  \startpage{1}
  \received{..}
  \revised{..}
  \accepted{..}

  \journalyear{..}
  \journalvolume{..}
  \journalissue{..}

\maketitle
\section{Introduction}
The low-latency anonymity network Tor protects the privacy of its users' internet browsing habits, allowing them to evade surveillance, tracking, and censorship. To do this, it encrypts internet traffic while routing it through a series of volunteer-run nodes, preventing any single node from knowing both the origin and destination of the communications \cite{Dingledine2004}. As a result, traffic destinations (such as websites) cannot determine the identities of their users, and local observers (such as malware, network administrators, or ISPs) cannot see the destinations of users' traffic. 

In the last decade, Tor's user base has increased to millions of daily users \cite{metrics}; over the same period, a series of papers have shown that Tor is vulnerable to a type of traffic analysis attack known as \textit{website fingerprinting} (WF) \cite{Panchenko2011, Cai2012, Wang2013, Hayes2015, Sirinam2018, Jansen2018, Rahman2019}. In a WF attack, a passive eavesdropper attempts to determine the destination of encrypted traffic by observing the sequences of packets sent and received by the user. The attack takes place between the user and the first node in the Tor network using features such as the volume of incoming and outgoing packets and the relative timing of packet bursts. With this information, the attacker then compares the collected packet trace to a database of website-trace pairs and classifies each trace accordingly. 

WF attacks are straightforward to carry out, as they require only passive eavesdropping from a local adversary, ISP, or Tor guard. Thus, WF attacks against Tor users represent a realistic threat and may be used to identify users who visit `censored' or forbidden websites. As this poses an obvious threat to users' privacy, researchers have developed a variety of defenses \cite{Wright2009, Luo2011, Cai2012, Cai2014, Juarez2015, Wang2017, Gong2020}. The goal of these defenses is to alter traffic in a manner that makes it difficult to determine which website is associated with each packet trace, and they typically operate by strategically adding `dummy' packets or by delaying packets, altering the patterns in each trace. 

However, the Tor Project has been hesitant to implement past defenses, as most would either impact user experience with increased latency, burden the Tor network with increased bandwidth, or require the creation and maintenance of additional infrastructure. Additionally, many of these defenses have been proven ineffective against the latest attacks, which utilize large data sets and sophisticated deep learning techniques. In response, we present RegulaTor, which provides strong protection against state-of-the-art WF attacks with moderate bandwidth overhead and a small latency penalty, but without requiring additional infrastructure or knowledge of other traces.  

Our key observation is that defenses that ``regularize'' traffic so that traces from different web pages are identical tend to be the most effective. However, these defenses are often the least efficient, as they incur high latency overhead in periods of heavy traffic and high bandwidth overhead in periods of light traffic. Still, constant rate traffic is \textit{just one of many} potential patterns for traffic regularization. Based on empirical evaluation of Tor web traffic, we find that there are common traffic patterns that avoid these traffic rate mismatches, allowing users to achieve the security benefits of regularization while greatly reducing the associated overhead. 

Accordingly, RegulaTor works by regularizing the size and shape of packet `surges' that frequently occur in download traffic, masking potentially revealing features. In this paper, `surge' is broadly defined as a large number of packets sent over a short period of time. To do this, whenever a download traffic `surge' arrives, RegulaTor starts sending packets at a set initial rate to avoid leaking information about the volume and length of the surge. Then, it decreases the packet sending based on a set `decay rate' parameter, which defines the shape of the surge. If no packets are available when one is scheduled, a dummy packet is sent instead. However, due to the heavy `burstiness' of web-browsing traffic, this download padding approach can be carried out with limited overhead. At the same time, RegulaTor sends upload packets at some fraction of the download packet sending rate. Moreover, sending upload packets based on download traffic usually incurs little latency overhead, as upload traffic mimics the download traffic (albeit with less volume) in web browsing traffic, as shown later in this paper. 

The RegulaTor approach deviates significantly from previously proposed WF defenses. First, it alters traffic in a time-sensitive manner with a focus on sending standardized surges, while other defenses (with FRONT \cite{Gong2020} a notable exception) tend to insert padding consistently along the packet sequence. Furthermore, it uses entirely different strategies to alter upload and download traffic, while other defenses pad traffic consistently regardless of direction. Lastly, RegulaTor uses the observed similarity between upload and download traffic to send upload traffic as a function of download traffic, preventing upload traffic from leaking any further information. 

We also re-evaluate previously presented attacks and defenses on our data set to enable direct comparison. In the closed-world setting with 95 websites, RegulaTor reduces the accuracy of the state-of-the-art attack, Tik-Tok, to only 25.4\% compared to 66.0\% for FRONT-2500. Furthermore, it requires a small latency overhead and a bandwidth overhead of only 79.7\%, while FRONT-2500 requires a bandwidth overhead of 119\%. In the open-world setting, RegulaTor's performance advantage is even more severe, reducing the $F_1$-score of a precision-tuned Tik-Tok attack to .135 compared to .625 for FRONT-2500. Thus, RegulaTor drastically outperforms comparable defenses in terms of efficiently defending traffic in the open-world setting.

\section{Preliminaries}
\begin{figure}
  \caption{WF Attack on Tor}
  \centering
  \includegraphics[scale=.7]{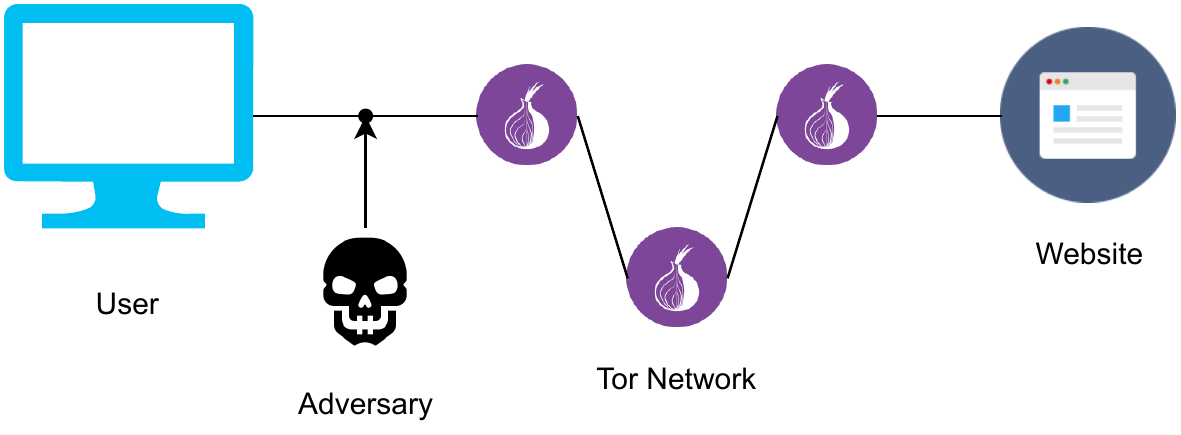}
  \label{fig:tor_wf}
\end{figure}

\subsection{Website Fingerprinting Background}
For WF attacks, the threat model is a passive, local adversary who eavesdrops on network traffic. Potential attackers include ISPs, network administrators, malware or packet sniffers, and the guard node of the Tor network, as shown in figure \ref{fig:tor_wf}. However, the attacker is not able to modify traffic in any way. To carry out the attack, the attacker first collects the traffic trace associated with the target's web browsing.  Then, the attacker compares the trace to a data set of (trace, web page) pairs, finding the matching web page based on a chosen set of features.

WF attacks are typically tested in two different settings: \textit{closed-world} and \textit{open-world}. In the closed-world setting, we assume that the target visits one of a defined set of web pages, and the attacker simply has to determine which of those web pages was visited. Because a real-world target may visit any number of web pages, this setting is unrealistic. Still, the closed-world accuracy is useful for comparing and evaluating defenses against a relatively strong WF attacker. 

On the other hand, the open-world setting provides a much more realistic setting where the target can visit any number of web pages. In this case, the attacker's goal is to determine whether or not the target visited a \textit{monitored} web page using information from the packet trace. While this task may be significantly more difficult given that the attacker cannot possibly train a model on all possible web pages, it is much more similar to the real-world scenario of an adversary trying to catch users visiting censored web pages.

Furthermore, we make several assumptions about the attacker's abilities originally stated by Juarez et al. \cite{Juarez2014}. The first assumption is that the attacker can detect the beginning and end of the page load, which is necessary to store a packet sequence representative of a given web page. Then, we also assume that all background traffic, which usually takes the form of other connections using the same Tor circuit, can be filtered out. Lastly, we assume that the attacker can create a data set representing the attacker's unique conditions. Some of these conditions include the Tor Browser Bundle (TBB) version, operating system, device hardware, and geographic location. Because these assumptions all favor the attacker, we expect that real-world defense performance is at least as high as demonstrated in this paper.

Hintz et al. \cite{Hintz2002} were the first to demonstrate success in WF attacks using the set of file transfer sizes to distinguish web pages. While this attack appeared to be effective against other privacy and anonymity systems, such as VPNs \cite{Shmatikov2006, Herrmann2009}, Tor at first appeared to be immune due to constant-size cell-padding, circuit multiplexing, and network-induced delays. However, later attacks utilized packet volume and timing features to greatly increase WF attack effectiveness against Tor \cite{Panchenko2011, Cai2012}. 

Wang et al. then made further improvements \cite{Wang2013}, first by improving data gathering and pre-processing, and later by using a large feature set with the k-nearest neighbors algorithm \cite{Wang2014a}. Later, the k-fingerprinting approach \cite{Hayes2015}, which used a series of simple but important features along with random forests and k-nearest neighbors, further improved accuracy. But, most importantly, k-fingerprinting demonstrated effectiveness in an open-world setting with a world size much larger than in previous works. Soon afterward, The CUMUL approach \cite{Panchenko2017} was presented by Panchenko et al. By sampling features from the cumulative representation of a trace, CUMUL outperforms previous attacks while staying computationally efficient.

While these models demonstrate high accuracy, the Deep Fingerprinting approach by Sirinam et al. \cite{Sirinam2018} uses convolutional neural networks to further improve the state-of-the-art in WF attack performance. Most importantly, Deep Fingerprinting manages to defeat the WTF-PAD defense and achieve high accuracy against the Walkie-Talkie defense in the closed-world scenario. As a result, this paper uses Deep Fingerprinting as a benchmark to test the RegulaTor defense and its rivals. The Tik-Tok attack \cite{Rahman2019} further improves the Deep Fingerprinting attack by using burst-level timing features. While previous attacks have ignored granular timing information, Tik-Tok uses the combination of directional and timing features to further improve performance against WF defenses, including the Walkie-Talkie defense. Accordingly, we use Tik-Tok as a benchmark in this paper as well. Other WF attacks and related techniques are discussed in the related works section.

Defenses evaluated in this paper include Tamaraw, WTF-PAD, and FRONT. Tamaraw \cite{Cai2014} serves as a baseline defense that regularizes traffic while providing proven security guarantees. However, it is not a practical defense, as it requires high bandwidth and latency overhead. WTF-PAD \cite{Juarez2014} uses adaptive padding to fill gaps in packet sequences to reduce the amount of information leaked by each trace. Alternatively, FRONT \cite{Gong2020} adds varying amounts of dummy packets to the beginning of packet sequences, making them more difficult to distinguish. Both WTF-PAD and FRONT require moderate bandwidth overheads and add no latency. Moreover, neither WTF-PAD nor FRONT require any information about other traces or additional infrastructure; thus, they are suitable comparisons for RegulaTor, which can also be implemented in a relatively straightforward manner. For a more in-depth summary of WF defenses, see the related works section.




\subsection{Metrics}
In the closed-world setting, the evaluation is straightforward, as the attacker simply needs to determine the correct website with the highest possible accuracy. However, the open-world setting is more complicated due to the imbalanced classes, as the attacker may only be able to monitor a small portion of web pages a user could visit. Thus, there is a strong potential for false positives to outnumber the true positive results. This possibility was discussed in the Tor community \cite{critique}, and attacks have since aimed to maximize the \textit{precision} of their classifiers. In the open-world scenario, the precision represents the monitored web pages that were detected compared to all of the web pages predicted to be in the monitored set. Accordingly, precision is a much more useful metric for evaluating the usefulness of a WF attack in a realistic scenario. Alternatively, recall, representing the fraction of monitored web pages that were retrieved, is also used in the open-world setting to demonstrate the sensitivity of an attack. An ideal attack has both high precision and high recall; however, one often comes at the expense of the other. 

\subsection{Defense Overhead}
WF defenses have the potential to impact user experience and increase strain on the Tor network by increasing latency, delaying page loading, and increasing the required bandwidth. As a result, we evaluate the \textit{latency overhead} and \textit{bandwidth overhead} of each defense.
We find latency overhead by calculating the additional time required to send the \textit{defended} trace compared to the undefended trace and dividing by the time required to send the original trace. In practice, we do this by subtracting the sending time of the last \textit{real} packet in the defended trace from the sending time of the last packet in the undefended trace. Intuitively, the latency overhead metric aims to indicate how much longer the user will have to wait to load a web page. 

The bandwidth overhead is found by dividing the number of dummy packets sent in the defended trace by the number of packets in the undefended trace. Essentially, it represents how much more data will be transmitted while loading a web page with the WF defense enabled. While increased bandwidth may strain the Tor network, discussion in the Tor community indicates that it is preferable to latency overhead \cite{padding}, which more directly impacts user experience. As a result, the RegulaTor defense aims to defeat WF attacks primarily through increased bandwidth, while increasing latency only minimally. 

\subsection{Data Sets}
Our primary evaluation uses two data sets provided by Sirinam et al. that were originally collected to test their Deep Fingerprinting attack \cite{Sirinam2018}. The closed-world data set was collected by visiting the homepages of the Alexa Top 100 sites 1,250 times each using tor-browser-crawler on ten low-end machines \cite{crawler}. The homepage visits were split into five batches where for each batch, each machine would access a website 25 times before moving on to the next website. Batching the crawling in this manner controls for both long and short-term variance, as described by Wang et al. \cite{Wang2013}. After discarding corrupted traces, 1000 instances of 95 sites were included in the final data set, which we refer to as DF-CW.

The open-world data set was collected by using tor-browser-crawler \cite{crawler} to visit the sites in the Alexa Top 50,000, excluding the top 100 sites crawled for the closed-world data set. Again, ten low-end machines were used, with each machine making one visit to the home pages of 5000 different sites. After discarding corrupted visits, 40,716 traces were included in the open-world data set, which we refer to as DF-OW.

To test the generalizability of the RegulaTor defense, we also use the cell traces provided by Wang et al. \cite{Wang2014a} to demonstrate their k-nearest neighbors attack as well as the recently collect Goodenough data set collect by Pulls \cite{pulls2020}. The former data set contains data for 100 websites with 90 instances each and was collected in 2014. The latter data set, created in 2020, collected 20 samples from each of 10 web pages for 50 websites, resulting in a total of 10,000 samples. For both data sets, only the closed-world traces are used for straightforward evaluation and comparison between defense settings. In this paper, we refer to the k-nearest neighbors data set as KNN and the Goodenough data set as GE. 

Furthermore, to test RegulaTor parameters in a real-world implementation, we use our pluggable transport implementation to collect 100 defended samples from each of the websites in the Alexa top 100. This data set was collected over one month in August 2021 and is referred to as PT. 

\section{RegulaTor}

\begin{figure}[t]
    \centering
    \includegraphics[scale=.5]{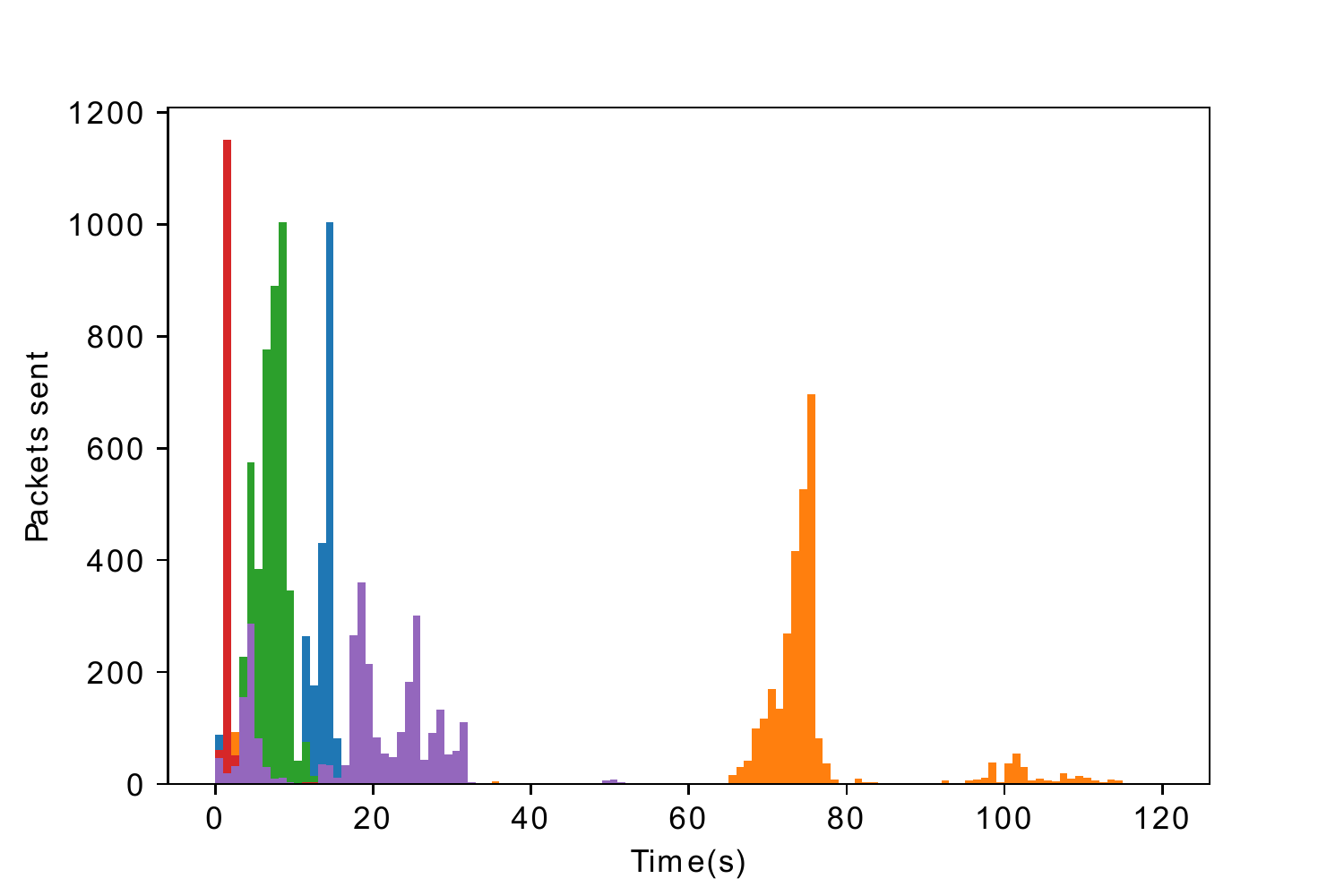}
    \captionsetup{justification=centering,margin=1cm}
    \caption{Examples of Tor download traffic during web page visits}
    \label{fig:bursty}
\end{figure}

\begin{figure}[t]
    \centering
    \includegraphics[scale=.5]{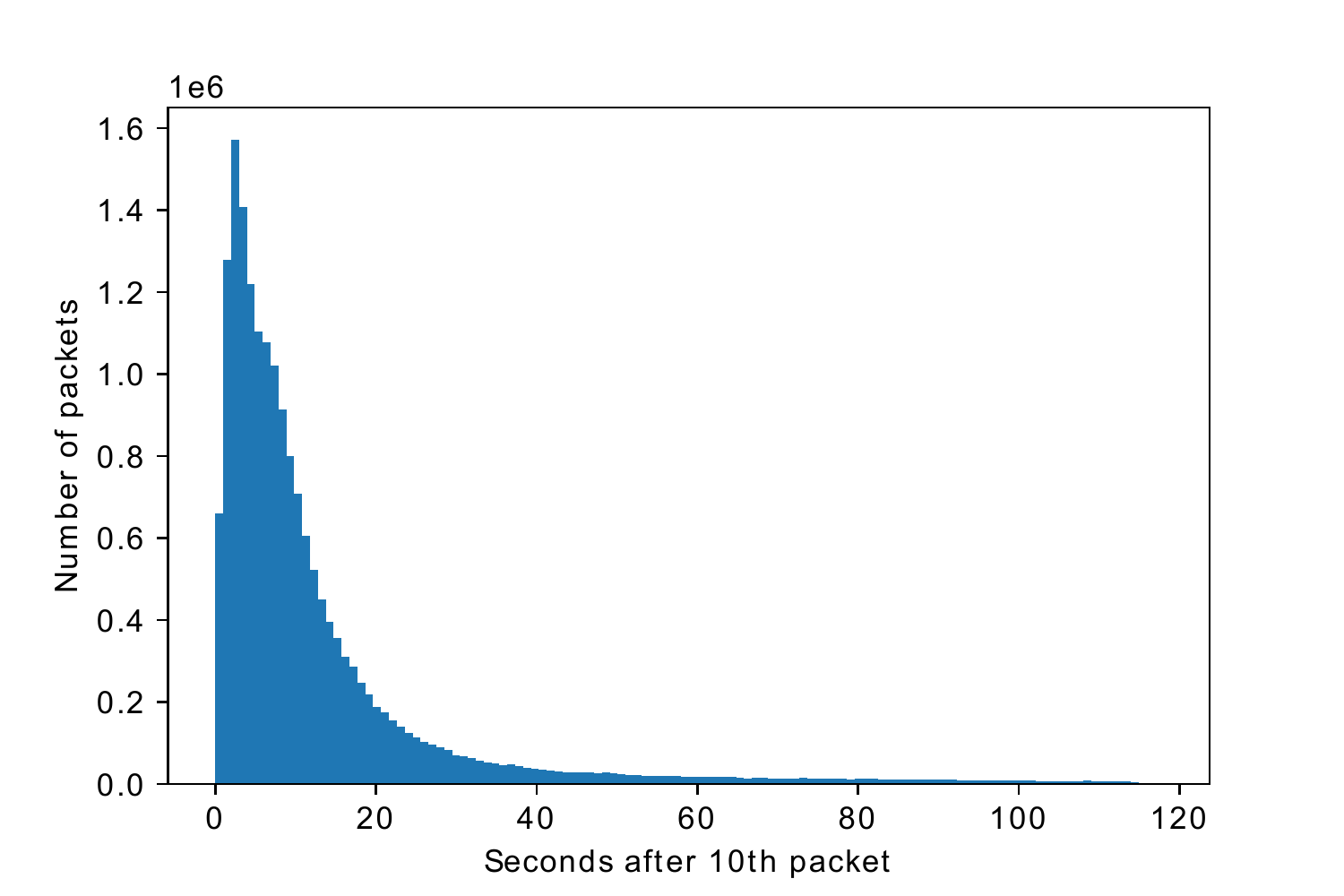}
    \captionsetup{justification=centering,margin=1cm}
    \caption{Decay of undefended download packet sending volume}
    \label{fig:packet_decay}
\end{figure}

\subsection{RegulaTor Justification}
\label{ssec:justification}

Given that the majority of relatively efficient previous WF defenses have been defeated, we aim to design a defense that can resist WF in both open-world and closed-world settings. Additional latency appears to be especially discouraged by Tor developers \cite{padding}, so we design RegulaTor to minimize the latency overhead of the defense. We also aim to keep the implementation simple and functional without requiring frequent tuning or collection of data sets, which is generally required in machine learning-based defenses. While this task is difficult, we find that there are common traffic patterns in Tor traffic that allow for an effective defense to operate with modest overhead.

First, we analyzed DF-CW and found that the \textbf{packet traces are particularly surge-heavy} in that they consist of infrequent and irregular `surges' of packets sent with little inter-packet delay. To illustrate this, we have plotted the download packet timing patterns from several randomly chosen traces (colored to distinguish between them) in figure \ref{fig:bursty}. Note that the traffic is characterized by occasional high-volume packet surges followed by periods of low traffic. We also find that, even though the average web page visit in DF-CW lasts about 28 seconds, the median interquartile range of packet times is only 3.96 seconds. This further demonstrates that the bulk of traffic is sent over a relatively short period of time. Furthermore, the location, size, and timing of these surges represent coarse features that leak a significant amount of information about the traces. To prevent these coarse features from leaking information about the traces, padding can be done in a more randomized manner to obfuscate the features, which was done in the FRONT defense; or, the surges in the packet sequences can be regularized in terms of size and location, which is the approach presented in this paper. 


Luckily, the packet sequences show that \textbf{surge patterns are often predictable}. To be specific, a majority of the packet sequences consist of an early sequence of upload packets followed by a sudden surge of download packets. While the download surge varies in terms of size and start time, it generally decays in volume soon after the initial spike as the web page finishes loading. This is shown in figure \ref{fig:packet_decay}, which represents the packet distribution for 10,000 randomly chosen traces. In order to control for the start time of the first surge, only traffic after the 10th packet is shown. Additionally, the median packet is sent 7.57 seconds after the 10th packet, further emphasizing how the bulk of traffic is sent soon after the beginning of a web page visit. Accordingly, regularizing the download packet sequences can be done efficiently by adding dummy packets to sequences with smaller initial surges and delaying packets in sequences with larger initial surges. 

\begin{figure}[b]
    \centering
    \includegraphics[scale=.5]{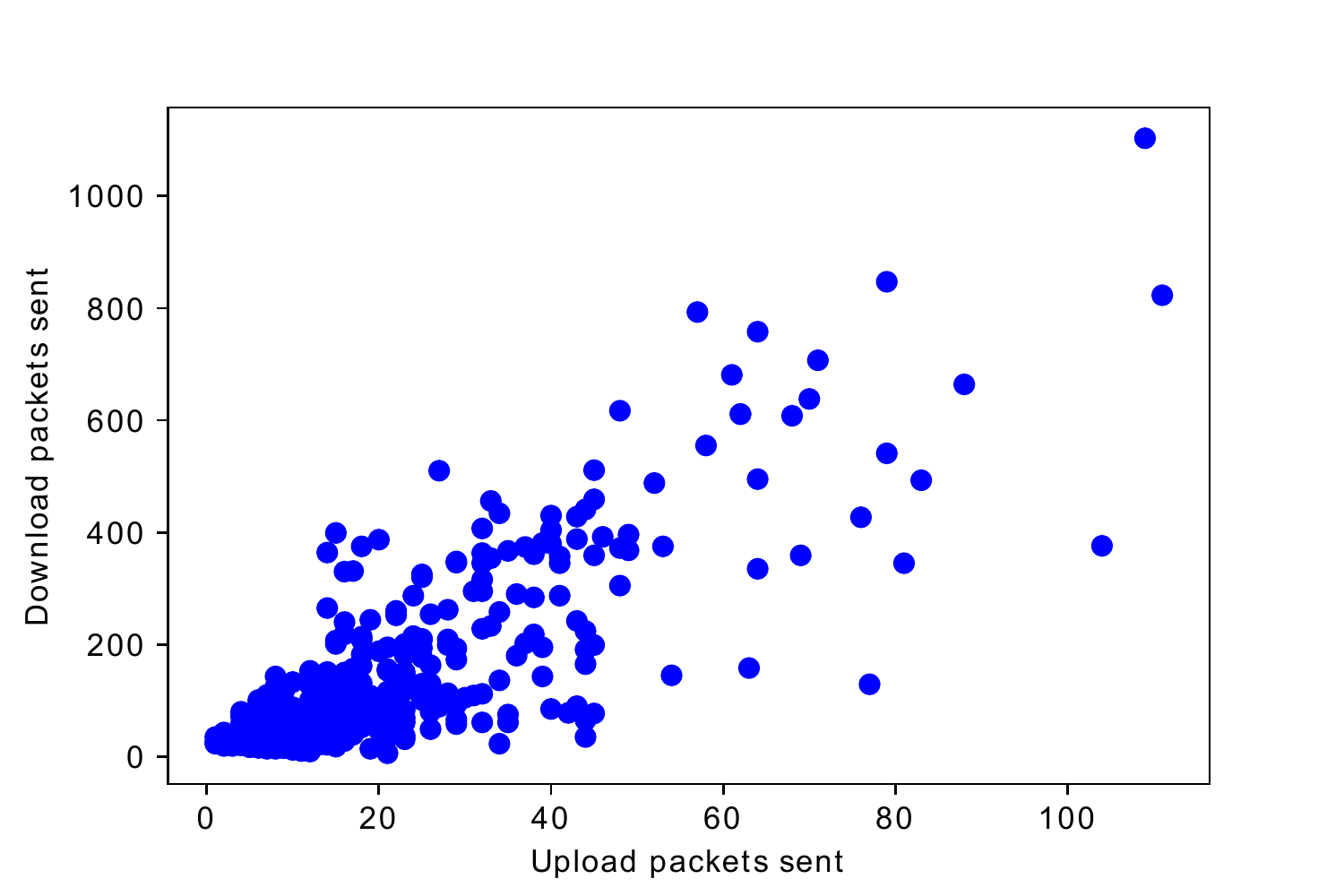}
    \captionsetup{justification=centering,margin=1cm}
    \caption{Download vs. upload traffic for each second}
    \label{fig:traffic_scatter}
\end{figure}

Furthermore, the \textbf{timing of upload packets imitates the timing of download packets}, despite the relatively low volume of upload traffic, as outgoing requests are generally quickly responded to by the web page. The correlation between upload and download traffic volume is illustrated in figure \ref{fig:traffic_scatter}, where the traffic from 30 randomly chosen traces is split into 1-second bins and plotted based on the number of upload and download packets sent in that period. Most importantly, figure \ref{fig:traffic_scatter} shows that if a substantial number of download packets are being sent, then upload packets are being sent concurrently. This presents an opportunity for RegulaTor to defend the upload packet sequence as well: by modeling upload packet sending as a function of the download packet sequence, the upload traffic leaks no further information about the destination web page. Additionally, RegulaTor can minimize upload packet latency increases by intentionally overestimating the upload sending rate, which is relatively efficient given the lower volume of upload packets.

\begin{figure}[b]
    \centering
    \includegraphics[scale=.5]{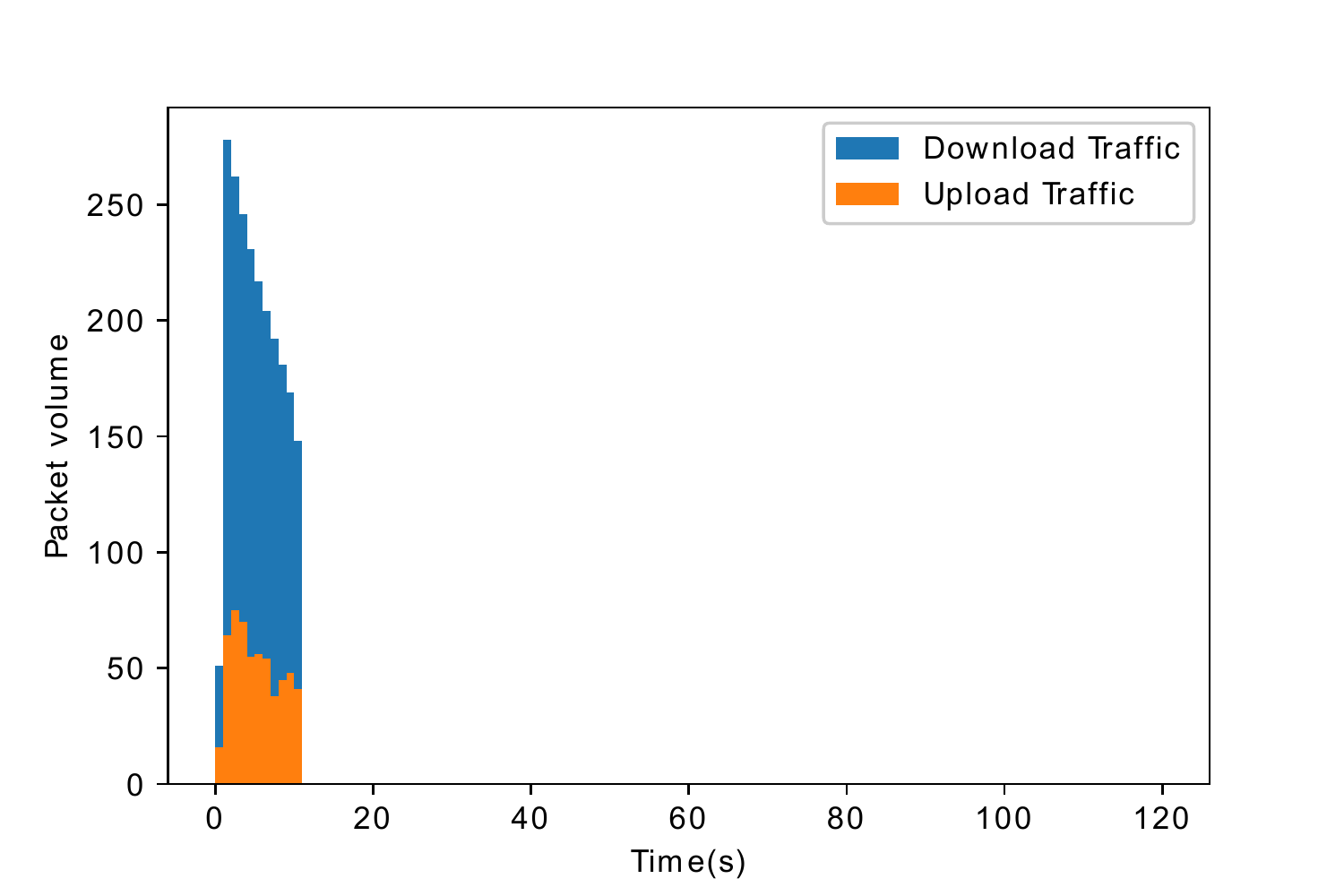}
    \captionsetup{justification=centering,margin=1cm}
    \caption{RegulaTor-defended Trace}
    \label{fig:def_trace_ex}
\end{figure}

For a visual representation of the RegulaTor defense, see figure \ref{fig:def_trace_ex}, which illustrates the rate of download and upload packet sending at different points in time for a single defended web page visit. 

\subsection{RegulaTor Design}

\begin{table}
\centering
\begin{tabular}{c|c}
\hline
    
     Parameters & Description  \\
     \hline
     \hline
     R & Initial surge rate\\
     \hline
     D & Packet sending decay rate\\
     \hline
     T & Surge threshold\\
     \hline
     N & Padding budget\\
     \hline
     U & Download-upload packet ratio\\
     \hline
     C & Delay cap\\
\end{tabular}
\vspace{5pt}
    \captionsetup{justification=centering,margin=2cm}
    \caption{RegulaTor parameter descriptions}
    \label{tab:regulator_parameters}
\end{table}

\begin{algorithm}
\caption{RegulaTor download padding main loop}
\begin{algorithmic}
\WHILE{$<$ 10 packets scheduled}
\STATE wait
\ENDWHILE
\STATE
\STATE $surge\mhyphen time \leftarrow$ {\small CURRENT-TIME}
\STATE $next\mhyphen packet \mhyphen time \leftarrow$ {\small CURRENT-TIME}
\WHILE{web page downloading}
\IF {{\small CURRENT-TIME} $\ge next\mhyphen packet \mhyphen time$}
\STATE
\STATE $target\mhyphen rate \leftarrow RD^{(\textnormal{CURRENT-TIME} - surge\mhyphen time)}$
\STATE

\IF {$target\mhyphen rate < 1$}
\STATE $target\mhyphen rate \leftarrow 1$
\ENDIF
\STATE

\IF{$waiting\mhyphen packets > T\cdot target\mhyphen rate$}
\STATE $surge\mhyphen time \leftarrow current\mhyphen time$
\ENDIF
\STATE

\IF{\small{NUM-WAITING-PACKETS} $= 0$}
\IF {$sent\mhyphen dummy\mhyphen packets < N$}
\STATE \small{SEND-DUMMY-PACKET}
\STATE $sent\mhyphen dummy\mhyphen packets \leftarrow sent\mhyphen dummy\mhyphen packets + 1$

\ENDIF
\ELSE
\STATE \small{SEND-PACKET}
\ENDIF
\STATE
\STATE $time\mhyphen gap \leftarrow target\mhyphen rate^{-1}$
\STATE $next\mhyphen packet\mhyphen time \leftarrow next\mhyphen packet \mhyphen time + time\mhyphen gap$

\ENDIF
\ENDWHILE
\end{algorithmic}
\label{alg:download_padding_alg}
\end{algorithm}

To enact the defense, the client pads the upload packets, while the download padding can be carried out by a Tor bridge, middle node, or guard node. However, padding to the middle node is likely the most effective method, as a WF adversary may be located at the guard node. By only padding the traffic between the client and one of the first relays in the circuit, the real-world cost of the bandwidth overhead can be substantially minimized. In fact, Tor is generally bandwidth-limited by the relatively limited number of exit nodes \cite{Juarez2015}, further minimizing the impact that RegulaTor-defended traffic will have on the network. 

The RegulaTor defense pads the download packets as follows: the first download packets are sent at a constant rate until 10 packets have been sent. This is done to avoid sending the initial RegulaTor surge before the surge of original data has been scheduled while also allowing the circuit-building and TLS handshake to finish. At this point, RegulaTor begins to send a surge of packets at the initial surge rate, $R$, which represents the number of packets sent per second. However, the sending rate is reduced according to the decay constant, $D$, such that the sending rate is $R^{Dt}$, where $t$ is the number of seconds since the surge began. Afterward, if the original packet sequence again calls for a significant number of packets to be sent and the queue of waiting packets increases to some threshold, then RegulaTor sends another surge of download packets to minimize the potential for increased latency. This threshold is calculated as the surge threshold, $T$, multiplied by the target rate.

To minimize additional bandwidth overhead, the defense draws a random padding budget from $(0, N)$, where $N$ is the maximum padding size. Once $N$ dummy packets have been sent, the sending of dummy packets is stopped, though real packets may still be delayed. The random choice of $N$ also functions to vary the total volume of the packet sequences, which reduces the amount of information leaked by the volume of the sequence.

RegulaTor pads the upload trace at a constant rate until the initial download surge begins to arrive to accommodate the initial web page request. At this point, the client schedules upload packets to send at some ratio, $U$, of the download packet sending rate (e.g. the client will send upload packets at 1/3 the rate of the download packets for $U = 3$). $U$ is chosen to minimize upload packet sending latency, so it will typically send a significant number of dummy packets; however, this is not costly bandwidth-wise, as the upload packet sequences are typically small. Additionally, if any upload packets have been delayed for more than $C$ seconds, then they are sent immediately to prevent excessive latency. 

Algorithm \ref{alg:download_padding_alg} demonstrates how the download packing sending schedule is determined, and table \ref{tab:regulator_parameters} contains the relevant parameters and their descriptions. 

\subsection{Parameter Tuning}

To determine the specific parameter values for RegulaTor, we used the Tree-Structured Parzen Estimator (TPE) technique, which is usually used for hyper-parameter optimization for learning algorithms \cite{tpe}. TPE creates models of hyper-parameter performance based on previous trials while determining which hyper-parameters to use next. While RegulaTor is not a learning algorithm, we are still able to use the TPE approach to determine high-performing parameter combinations for the defense. 

To implement parameter tuning, we used the Python library hyperopt \cite{hyperopt} and determined RegulaTor's performance by simulating RegulaTor on DF-CW and testing the performance of Tik-Tok \cite{Rahman2019} against the RegulaTor-defended data set. We chose Tik-Tok as the WF attack so that RegulaTor would be tuned to avoid leaking timing information, which Tik-Tok can detect with considerable effectiveness. 

Then, to determine the performance of parameter combinations, we created a loss function based on a weighted combination of the latency overhead, bandwidth overhead, and WF attack accuracy. This also allows us to alter RegulaTor based on which properties (e.g. low latency) are most desired. In our evaluation, we present two RegulaTor defenses: one that emphasizes higher performance and bandwidth overhead (RegulaTor-Heavy), and another that aims to achieve moderate performance with a lower bandwidth overhead (RegulaTor-Light). 

\section{Defense Evaluation}

\begin{table*}[]
    \centering
    \begin{tabular}{p{2cm}||p{7cm}}

    \hline
    Defenses & Parameters\\
    \hline
    \hline
    Tamaraw & $\rho_{out}=.04, \rho_{in}=.012, L=100$\\
    WTF-PAD & normal\_rcv\\
    FRONT-1700 & $N_s = N_c = 1700, W_{min}=1, W_{max}=14$\\
    FRONT-2500 & $N_s = N_c = 2500, W_{min}=1, W_{max}=14$\\
    RegulaTor-Light & $R=260, D=.860, T=3.75, N=2080, U=4.02, C=2.08$\\
    RegulaTor-Heavy & $R=277, D=.940, T=3.55, N=3550, U=3.95, C=1.77$\\
    \hline

\end{tabular}
    \vspace{5pt}
    \captionsetup{justification=centering,margin=2cm}
    \caption{Defense Parameters}
    \label{tab:parameters}
\end{table*}
In this section, we evaluate RegulaTor and several other WF defenses in terms of their closed-world performance, open-world performance, and overhead. Table \ref{tab:parameters} presents the parameters used in each of the examined defenses. For defenses other than RegulaTor, default parameters were used as provided by their authors \cite{Cai2014, Juarez2015, Gong2020}

For evaluation, we simulated each defense on the undefended data sets DF-CW and DF-OW to create defended data sets for the closed-world and open-world settings. To simulate WTF-PAD and FRONT, we used the code provided by their authors. For Tamaraw, we used Tao Wang's implementation \cite{tamarawcode}. Then, we re-trained each WF attack on the defended data sets. The attacks used in this section used default parameters as well, except for CUMUL, which performs SVM hyperparameter tuning. 

To demonstrate RegulaTor's generalizability, we also simulate RegulaTor on data sets used in past works, present the performance of a real-world RegulaTor implementation, and investigate parameter trade-offs. Then, we discuss parameter stability over time and the practicality of RegulaTor deployment. 

\subsection{Closed-World}

\begin{table*}[]
    \centering
    \begin{tabular}{p{2.5cm}||p{2cm}|p{2cm}|p{1.5cm}}

    \hline
    Defenses & Tik-Tok & DF & CUMUL\\
    \hline
    \hline
    Undefended & 97.0\% & 98.4\% & 97.2\%\\
    WTF-PAD & 94.2\% & 92.4\% & 59.4\%\\
    FRONT-1700 & 78.2\% & 77.5\% & 31.6\%\\
    FRONT-2500 & 66.0\% & 69.8\% & 17.1\%\\
    RegulaTor-Light & 34.8\% & 23.3\% & 20.8\%\\
    RegulaTor-Heavy & 25.4\% & 19.6\% & 16.3\%\\
    Tamaraw & 10.1\% & 9.9\% & 17.0\%\\
    \hline

\end{tabular}
    \vspace{5pt}
    \captionsetup{justification=centering,margin=2cm}
    \caption{Closed-World Accuracy}
    \label{tab:defenses}
\end{table*}

Table \ref{tab:defenses} presents the accuracy achieved by WF attacks on the examined defenses. Since the experiment was done in the closed-world setting, accuracy is the only metric presented, and the DF-CW data set was used.
Each attack achieved high accuracy on the undefended data set with Deep Fingerprinting achieving the highest accuracy. Tik-Tok and Deep Fingerprinting were highly effective against WTF-PAD, demonstrated moderate effectiveness against FRONT defenses, and were only marginally effective against RegulaTor defenses. However, CUMUL accuracy sharply decreased for all defenses. No attack was effective against Tamaraw, which is included as an example of a defense with strong theoretical foundations but impractically high overhead. 

Tik-Tok outperformed Deep Fingerprinting against the RegulaTor defense, likely due to the use of packet timing to provide further information. Given that RegulaTor generally sends upload packets at regular intervals, attacks that represent the traces using only packet direction, such as Deep Fingerprinting, are unable to achieve high accuracy. The usefulness of timing information was most apparent with RegulaTor-Light, which reduced Tik-Tok accuracy to 34.8\% compared to 26.2\% for Deep Fingerprinting. 

Both RegulaTor-Heavy and RegulaTor-Light reduced Tik-Tok and Deep Fingerprinting accuracy significantly more than any other practical defense. Even when comparing the `light' version of RegulaTor to the bandwidth-heavy version of FRONT, Tik-Tok accuracy is decreased from 66.0\% to 34.8\% and Deep Fingerprinting accuracy is decreased from 69.8\% to 23.3\%. However, CUMUL accuracy is not necessarily decreased when comparing FRONT to RegulaTor-Light. We suspect that this is because CUMUL derives a majority of its features from the cumulative representation of the trace, and FRONT effectively obfuscates early incoming and outgoing bursts in the trace, modifying the cumulative representation substantially. 

Furthermore, RegulaTor-Light manages to outperform its rivals while incurring a small latency overhead and a bandwidth overhead similar to that of WTF-PAD. RegulaTor-Heavy then further increases this margin with a bandwidth overhead still less than that of the lower-overhead version of FRONT.

\subsection{Open-world}
\subsubsection{Open-World Setup}

While the closed-world setting is useful for illustrating the relative effectiveness of WF attacks and defenses, it is not a particularly strong indicator of real-world usefulness. As described earlier, the open-world model is characterized by a user who can visit any web page and an attacker who monitors a subset of those web pages while training a classifier to determine whether the visited web page is in that monitored set. This attack is typically more difficult to carry out, given that the attacker cannot train the model on many of the packet sequences in the unmonitored set. 

In this experiment, when the attacker determines that a packet sequence represents a visit to a monitored web page, this prediction is a true positive if correct and a false positive otherwise. Similarly, if the attacker determines that a packet sequence represents a visit to an unmonitored web page, it is a true negative if correct and a false negative otherwise. An attacker is said to have determined that a web page is in the monitored set if the attacker's output probability is above a certain threshold. By varying this threshold, we can calibrate the attacks to achieve high recall or high precision. 

However, true positive rate (TPR) and false positive rate (FPR) are not relevant measures of model performance, since the set of unmonitored web pages may be much larger than the set of monitored ones, causing heavily unbalanced classes. Furthermore, as the unmonitored class size grows, the number of false positives may begin to rival or surpass the number of true positives, making WF attacks impractical \cite{Juarez2014, critique}. Thus, as described in previous discussions of WF open-world metrics \cite{Juarez2014, Juarez2015, Panchenko2017, Sirinam2018, Gong2020}, we instead use precision-recall curves to evaluate WF attacks and defenses.

To carry out the open-world experiment, we use the two strongest WF attacks (Tik-Tok and Deep Fingerprinting) and evaluate them against the defenses used in the closed-world setting (except for Tamaraw, which prevented both attacks from reporting more than negligible true positives for many thresholds used). Our training and testing setup models that of the open-world experiment in Deep Fingerprinting \cite{Sirinam2018}, which used 85,500 monitored traces (900 each from 95 web pages) with 20,000 unmonitored traces in the training set and 9500 monitored traces (100 each from 95 web pages) with 20,000 unmonitored traces in the testing set. 

\subsubsection{Open-World Results}

\begin{figure}
    \centering
    \includegraphics[scale=.5]{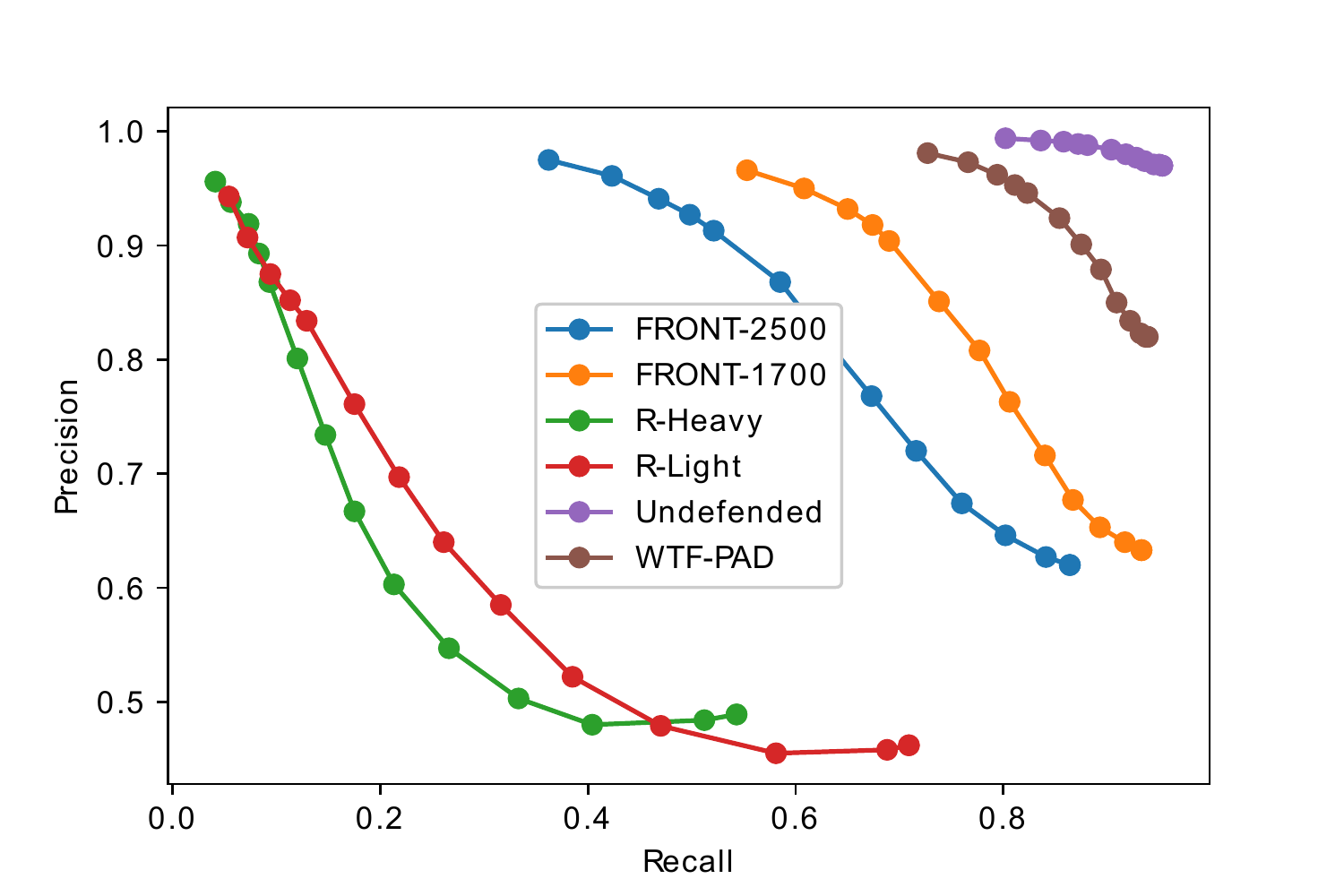}
    \captionsetup{justification=centering,margin=2cm}
    \caption{Tik-Tok Precision-Recall}
    \label{fig:ttpr}
\end{figure}

\begin{figure}[b]
    \centering
    \includegraphics[scale=.5]{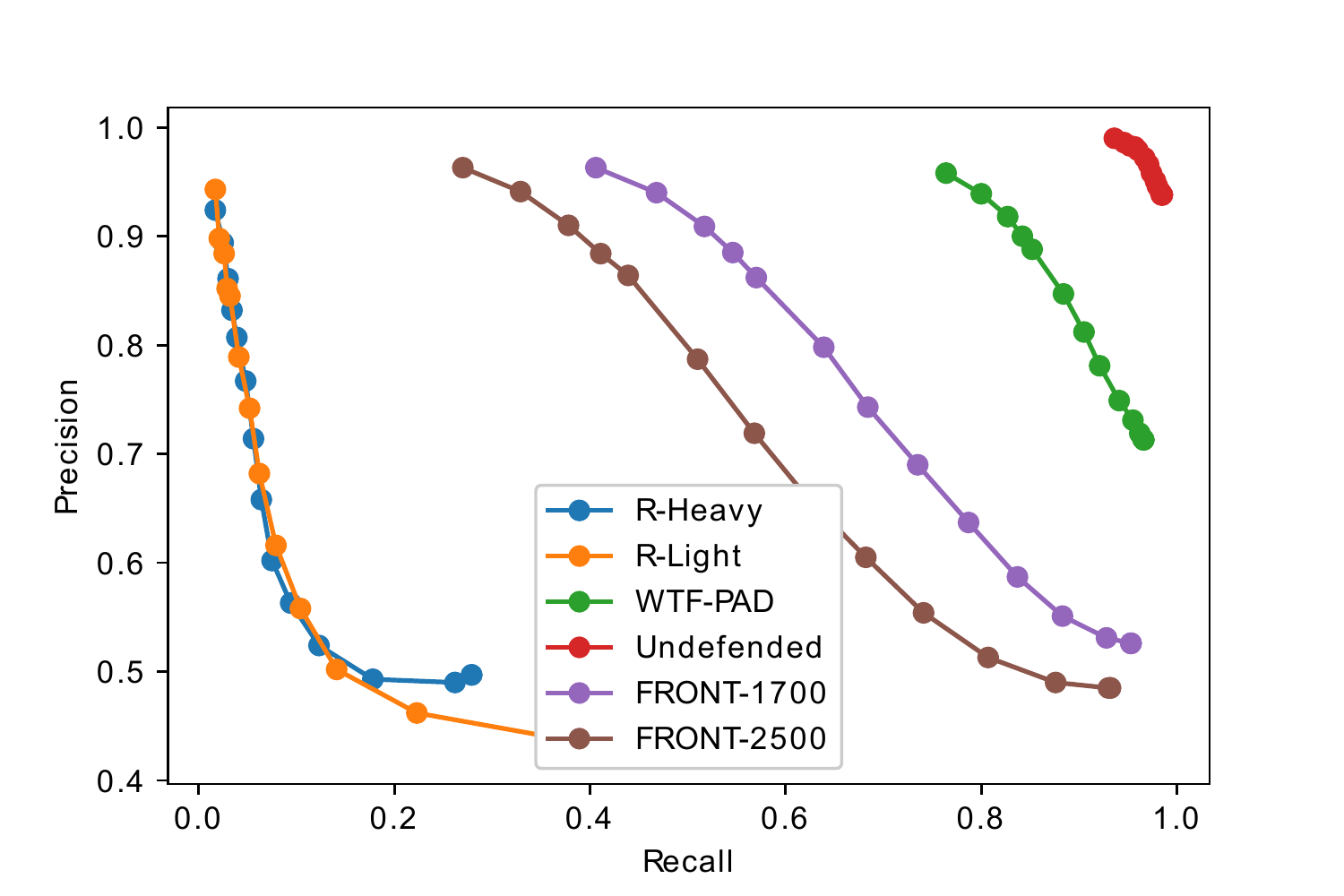}
    \captionsetup{justification=centering,margin=2cm}
    \caption{Deep Fingerprinting Precision-Recall}
    \label{fig:dfpr}
\end{figure}

The precision-recall curves for the Tik-Tok attack against the WF Defenses are shown in figure \ref{fig:ttpr}. As expected, Tik-Tok achieves both high precision and high recall in the undefended data set, demonstrating the effectiveness of the attack in the open-world setting. 

WTF-PAD and the FRONT defenses, alternatively, manage to drive down the precision-recall curve, forcing the attacker to choose between high precision with moderate recall and vice versa. However, FRONT-1700 does this more effectively than WTF-PAD, and FRONT-2500 further increases defense performance. 

As shown, RegulaTor defenses greatly outperform their rivals, allowing for high precision only with very low recall. As a result, the Tik-Tok attack is not particularly useful, as few packet sequences can be determined to be in the monitored set with high reliability.

The results for the open-world Deep Fingerprinting attack, shown in figure \ref{fig:dfpr}, are similar, except that each of the defenses is moderately more effective in terms of driving down precision and recall values, and the RegulaTor defense instances are about equal in terms of performance. The similarity between the RegulaTor defenses is likely the result of RegulaTor's ability to mask most features based on directional information.

To better compare the open-world performance of the tested defenses, consider that the $F_1$-scores of a precision-tuned Tik-Tok are $.870$, $.625$, and $.135$ against WTF-PAD, FRONT-2500, and RegulaTor-Heavy respectively. Note that a lower $F_1$-score implies that the associated defense is more effective. The relatively small $F_1$-score associated with RegulaTor further demonstrates its ability to prevent Tik-Tok from detecting a substantial number of monitored web pages without a high false positive rate as well.


Essentially, the open-world results indicate that the RegulaTor defenses prevent the studied WF attacks from achieving high degrees of precision and recall in a realistic setting. This is made even more apparent by the fact that the open-world setup used in this paper favored the attacker by using a test set where 9500 of the 29,500 traces were from the monitored class. In the real-world setting, it is unlikely that this proportion of all web pages would be monitored. 

\subsection{Alternate Data Sets}
\label{ssec:alt}
To confirm that the RegulaTor defense generalizes beyond a given data set, we test its performance on two other publicly available website fingerprinting data sets. For simplicity, we only evaluate Regulator-Heavy against the most effective WF attack (Tik-Tok) in the closed-world setting.  

First, we test whether the parameters determined from tuning on one data set provide effective defense for other data sets. To do this, we simulated the Regulator-Heavy defense setting on KNN and tested its performance against Tik-Tok. We found that it performed well, reducing Tik-Tok accuracy to 17.8\% with 5.1\% latency overhead and 77.3\% bandwidth overhead. For comparison, Front-2500 defends KNN with a Tik-Tok accuracy of 44.9\% and 98.3\% bandwidth overhead. In both cases, slightly reduced accuracy is expected given that there are only 90 instances of each website compared to 1000 in DF-CW, limiting the data available to Tik-Tok. Still, it appears that RegulaTor parameters can generalize across data sets, especially given that the data sets were collected a few years apart. 

However, it is important to note that the ideal sizes of the packet surges in the RegulaTor defense are dependent on the volume of the undefended traffic sequence. If the RegulaTor surges are too small, then latency will unnecessarily increase as data waits to be sent. While DF-CW and KNN contained similar traffic volume, we may not be able to make this assumption in the real-world setting. To demonstrate this situation, we simulate RegulaTor-Heavy on GE, which contains much higher traffic volume at 5663.9 packets per trace compared to 2100.9 for DF-CW and 1807.6 for KNN. After simulating RegulaTor on GE, we find that RegulaTor-Heavy reduces Tik-Tok to an attack accuracy of 11.3\% with 15.1\% latency overhead and 39.6\% bandwidth overhead. Here, the defense performance is high, but latency overhead is higher as well.

To adjust for the volume of traffic in the data set, we can increase the initial surge rate and padding budget of the defense proportionally to the increased volume of the traffic in the target data set. While this implies that some data collection should be occasionally done to implement the defense, this collection can be fairly minimal, as the adjustment only needs a rough estimate of relative traffic volume. 

To demonstrate this adjustment using GE, we multiply the initial surge rate by 2.431, as this represents a relative increase of traffic volume in GE. Additionally, we increase the padding budget by a similar amount. This results in an altered Regulator-Heavy defense with an initial surge rate of 673 and a padding budget of 8030. When simulated on GE, Tik-Tok accuracy is only 5.2\%, defense latency overhead is 2.9\%, and bandwidth overhead is $82.9\%$. So, while bandwidth is increased, the altered defense is significantly more effective and operates with reduced latency overhead. For comparison, we test a FRONT defense with increased dummy packet volume to match the FRONT-2500 bandwidth overhead in the original paper \cite{Gong2020}. Using a padding budget of 2830 on both the upload and download packets, we find that the FRONT defense reduces Tik-Tok accuracy to 43.4\% with a bandwidth overhead of 45.8\%. Thus, adjusted RegulaTor is still effective relative to FRONT. 

In summary, RegulaTor-Heavy remains effective even when tuned on one data set and then used on another. However, differences in traffic volume between the data set used for tuning and the target data set may cause RegulaTor to incur excess latency or bandwidth overhead. As a result, the initial surge rate and padding budget should be proportionally adjusted as described previously. Then, RegulaTor-Heavy remains highly effective while incurring the intended bandwidth and latency overhead. 

\subsection{Real-world Performance}

To obtain more accurate overhead estimates and confirm that RegulaTor could be smoothly used with Tor, we implemented the RegulaTor defense as a pluggable transport \cite{pt}. Pluggable Transports (PTs) transform traffic between the client and a bridge in order to disguise the Tor traffic and prevent censorship. Our specific approach was to host a Tor bridge and use the WFPadTools framework \cite{wfpadtools}, which is based on the Obfsproxy pluggable transport \cite{obfsproxy}, to build the RegulaTor defense. Additionally, we used a modified version of tor-browser-crawler \cite{Juarez2014, Rahman2019} to collect the traces for both the RegulaTor-Heavy defense and a `dummy' transport for comparison. The parameters used in the pluggable transport version of RegulaTor were based on the RegulaTor-Heavy parameters, except that the initial surge rate and download-upload packet ratio parameters were adjusted based on the observed traffic patterns, as described in section \ref{ssec:alt}. 

Then, we used our pluggable transport implementation to collect a RegulaTor-defended data set, PT, consisting of 100 websites and 100 samples per website. To determine the initial surge rate and padding budget, we first collected 10 samples from each of the websites and calculated that the average trace length was 2697.2. Comparing this to the traffic volume found in DF-CW, which was used to tune the original Regulator-Heavy, we then proportionally increased the initial surge rate to 356 and the padding budget to 4564. The remaining parameters were left unchanged, as they appear to generalize regardless of traffic volume. 

Using Tik-Tok to test the closed-world WF attack on PT, we record an accuracy of $11.6\%$. Compared to the traces collected using a dummy pluggable transport, the adjusted RegulaTor-Heavy defense operates with a latency overhead of 13.9\% and a bandwidth overhead of 78.2\%. While the observed latency overhead appears somewhat higher than our original prediction (as discussed in section \ref{ssec:overhead}), the adjusted RegulaTor-Heavy defense is more effective as expected. These results demonstrate that RegulaTor can be effective on live traffic and that the parameters found from tuning on a previously collected data set are still valid for a real-world implementation. 

\subsection{Overhead}
\label{ssec:overhead}
\begin{table}[]
    \centering
    \begin{tabular}{p{2cm}||p{1.75cm}|p{2.25cm}}

    \hline
    Defenses & Latency OH & Bandwidth OH\\
    \hline
    Tamaraw & 36.9\% & 196\%\\
    WTF-PAD & 0\% & 54.0\%\\
    FRONT-1700 & 0\% & 81.0\%\\
    FRONT-2500 & 0\% & 119.0\%\\
    RegulaTor-Light & 8.9\% & 48.3\%\\
    RegulaTor-Heavy & 6.6\% & 79.7\%\\
    \hline

\end{tabular}
    \vspace{5pt}
    \captionsetup{justification=centering,margin=2cm}
    \caption{Defense Overheads on DF-CW}
    \label{tab:overhead}
\end{table}

Table \ref{tab:overhead} summarizes the bandwidth and estimated latency overheads for each of the tested defenses on DF-CW. The WTF-PAD and FRONT defenses operate with no additional latency, while the RegulaTor defenses incur a small delay on some packet sequences, and the Tamaraw defense causes a substantial delay. Still, since this paper measures latency overhead as the delay of the last `real' packet, rather than the last dummy packet sent, Tamaraw's latency overhead may appear smaller than in previous works. 
Here, RegulaTor's latency overhead is an estimate based on the sum of the delay of the last real download packet and the maximum delay of any upload packet. The maximum delay of any upload packet is used because delayed upload requests may delay requests to the web server, delaying download packets even more so. 

\begin{figure*}[]
    \centering
    \includegraphics[]{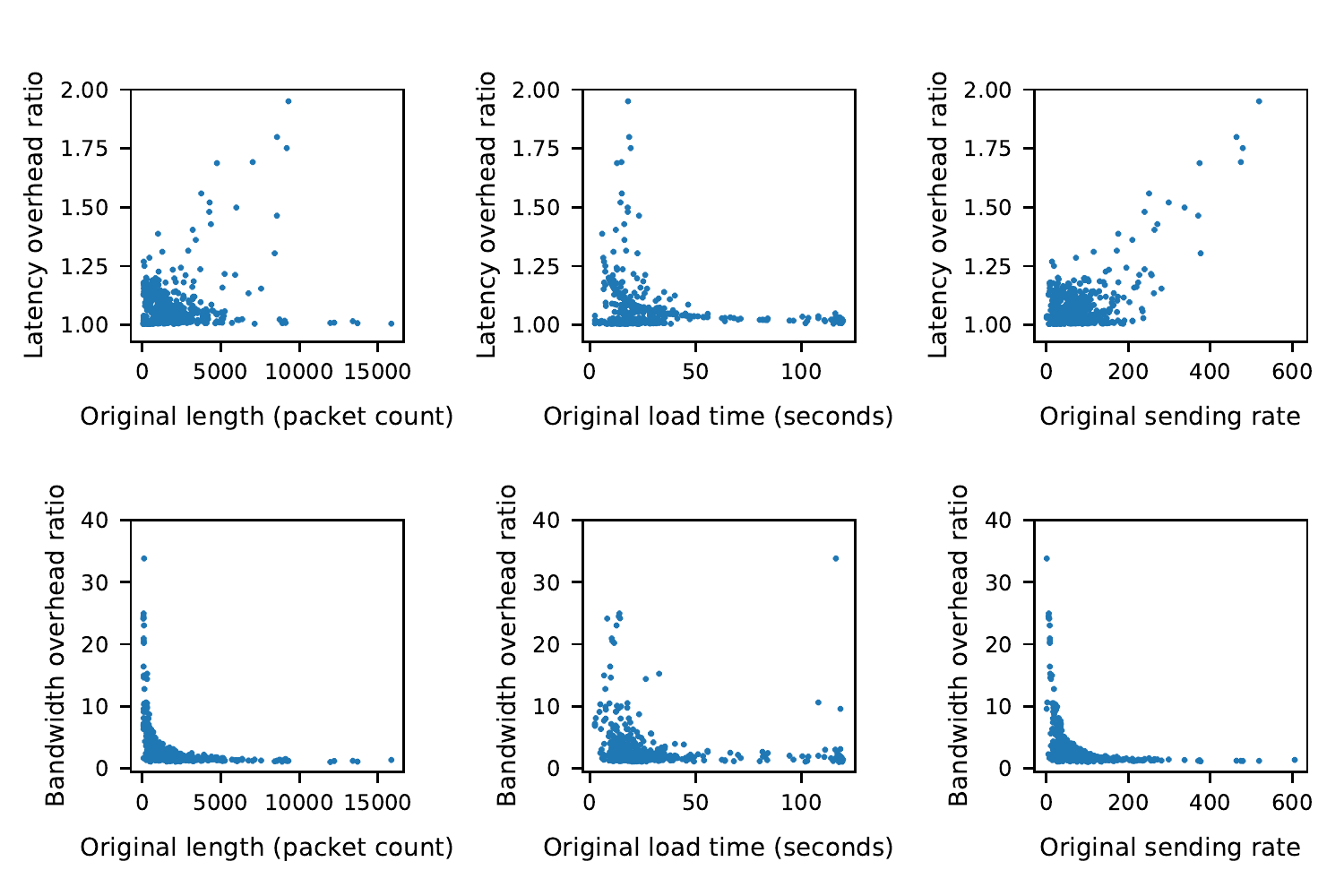}
    \captionsetup{justification=centering,margin=2cm}
    \caption{Overhead as a function of trace volume, load time, and packet sending rate}
    \label{fig:ls}
\end{figure*}

To illustrate how latency and bandwidth overhead are distributed across various websites, figure \ref{fig:ls} provides the bandwidth and latency multiples as functions of the original trace length, load time, and sending rate for 500 web sites randomly sampled from DF-CW. 

As expected, nearly all `short' traces and traces with a low sending rate have very low latency overhead. This is likely because RegulaTor schedules packets at a much higher rate than the original traces, delaying few packets. However, it is notable that packets with a very short load time often experience high latency overheads as well. This may be because traces with low loading times but high volume send packets at a high rate, which may exceed RegulaTor's sending rate. For the same reason, the sending rate of a trace appears to be correlated with latency overhead. Overall, a majority of traces incur little latency overhead, while traces with a high sending rate more often incur high latency overhead. 

Additionally, low trace length and sending rate correspond with increased bandwidth overhead. This is expected, as RegulaTor will likely schedule packets at a much faster rate, which results in the frequent sending of dummy packets. Accordingly, traces with high sending rates are associated with low bandwidth overhead ratio. Also, a moderate number of traces with low load time incur high bandwidth overhead. This negative relationship appears to be because traces with low loading time tend to have smaller packet counts, with occasional exceptions.

\subsubsection{Overhead-performance Trade-offs}

\begin{figure*}[h]
    \centering
    \includegraphics[scale=.95]{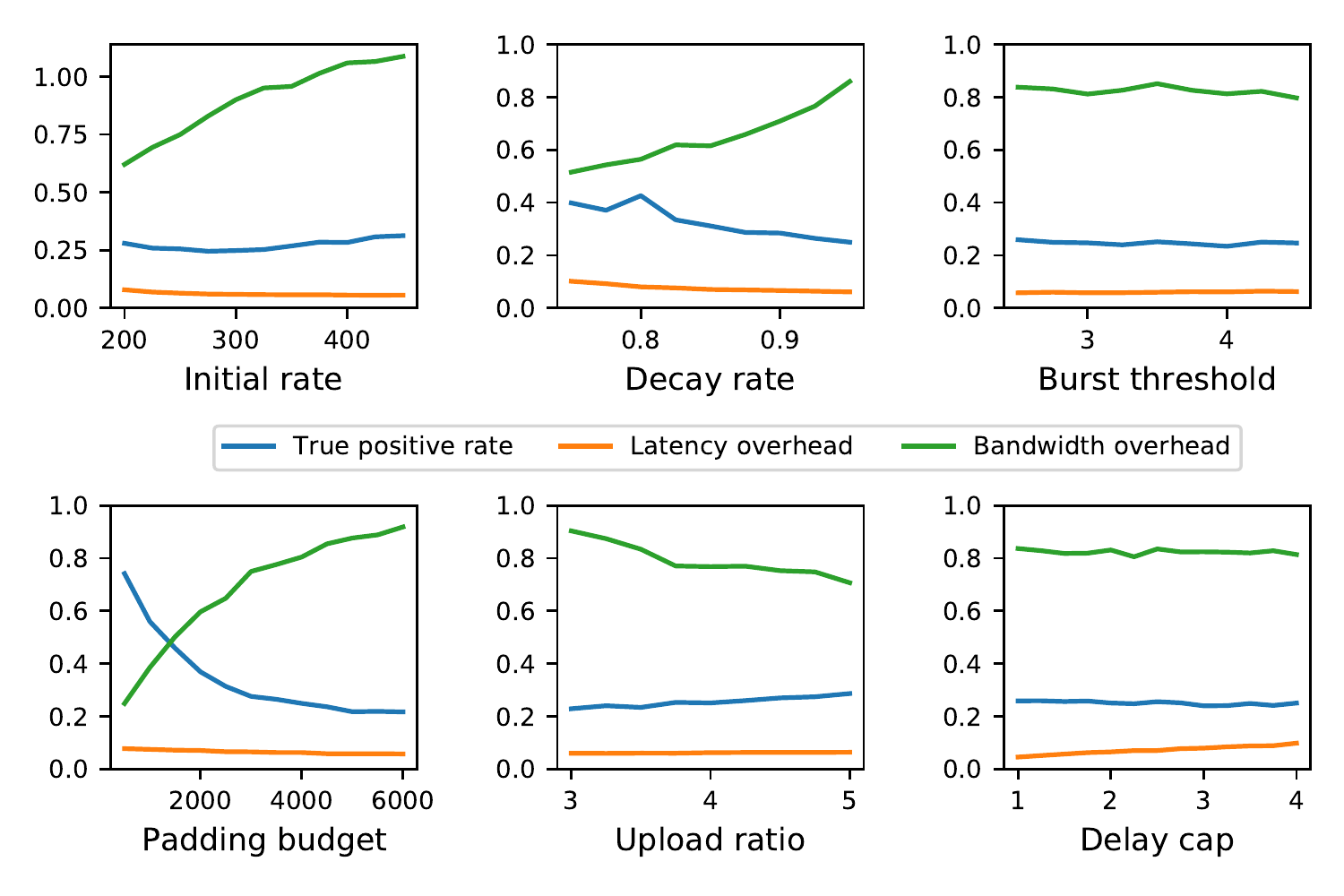}
    \captionsetup{justification=centering,margin=2cm}
    \caption{Overhead and closed-world performance by parameter value}
    \label{fig:ps}
\end{figure*}

To understand the impact of parameter choice, we also simulate RegulaTor on DF-CW with varied parameter values and record the latency overhead, bandwidth overhead, and Tik-Tok performance on the defended data set. To be specific, we used the RegulaTor-Heavy parameters and one by one varied the values of each parameter over a wide range. The results of these experiments are shown in figure \ref{fig:ps}. 

The experiments indicate that RegulaTor performance is fairly stable. Furthermore, varying the parameters allows for RegulaTor to choose different trade-offs between latency overhead, bandwidth overhead, and defense performance. Specifically, increasing the initial rate increases bandwidth and slightly decreases latency, while increasing the upload ratio decreases the bandwidth but weakens defense performance. Additionally, increasing the padding budget increases latency and defense performance while decreasing latency. 

Varying burst threshold and delay cap parameter values had little effect on bandwidth overhead and defense performance, though increasing delay cap increased latency overhead. Lastly, increasing decay rate improved defense performance and decreased latency at the expense of increased bandwidth overhead. 

\subsubsection{Real-world Latency Overhead}

While latency overhead can be estimated by simulating RegulaTor on Tor traces, the exact overhead is more difficult to determine, as one would have to accurately predict how the web server would respond to delayed upload packets. However, we can use the latency observed in the real-world data set, which was collected using a PT RegulaTor implementation. 

By comparing the loading time between the `dummy' PT and the RegulaTor PT, we find that the average web page loading time for the RegulaTor defense increased 13.9\%. While this exceeds the estimated latency for DF-CW, we find that this is partially explained by differences in traffic patterns. 

Specifically, the standard deviation of load time recorded by our dummy PT, at 42.3, is substantially larger than that of DF-CW, which was 28.0. As a result, the real-world RegulaTor implementation more frequently encounters web pages with short load time and high volume or web pages with particularly long loading times. In the former case, RegulaTor will likely delay loading, as it will send packets slower than the high rate in the original trace. In the latter case, RegulaTor will have significantly slowed packet sending near the end of the trace, which may then cause increased latency if a late surge of packets is sent. 

Even with somewhat increased latency, RegulaTor defense performance in the real-world is strong, reducing Tik-Tok accuracy to 11.6\%. Furthermore, if lower latency is desired, then parameters may be tuned to decrease it at the expense of a slight bandwidth or performance penalty. 



\subsection{Practicality}

Due to RegulaTor's simplicity, it appears that the parameters do not have to be frequently re-tuned over time. Specifically, as long as web traffic can be characterized as predictably `bursty' (as described in section \ref{ssec:justification}), then the surge threshold, delay cap, and decay rate parameters will remain stable. This is further supported by RegulaTor's effectiveness on various data sets with these parameters unchanged. Furthermore, the download-upload packet ratio will remain stable as long as the real-world ratio of download to upload traffic remains constant. While this is difficult to predict, it is notable that both DF-CW and the data collected with the dummy PT, which are collected with similar methodologies but several years apart, report a download-upload packet ratio of 5.96 and 6.66 respectively. Thus, we expect that the download-upload packet ratio parameter will remain somewhat stable. 

However, the initial surge rate and padding budget should be increased in response to changes in the average page load bandwidth. This increase appears likely to happen: according to the HTTP Archive \cite{archive}, the average resource transfer size of tracked URLs increased from 1412.1 KB on July 1, 2017 to 2148.7 KB on July 1, 2021. Accordingly, the ideal initial surge rate and padding budget for the RegulaTor defense may change over time; fortunately, this change is likely to be predictable and straightforward to determine, as one only has to crawl a set of web pages and determine the change in average traffic volume. 

In terms of the practicality of a real-world implementation, RegulaTor offers several advantages over previous defenses. First, RegulaTor's overhead is relatively moderate compared to existing regulation defenses, which tend to incur very high latency and bandwidth overhead. RegulaTor also operates without a database of traces associated with other websites, while defenses such as Glove \cite{Nithyanand2014}, Supersequence \cite{Wang2014a}, and Traffic Morphing \cite{Wright2009} require knowledge of other traces. Lastly, RegulaTor does not require additional infrastructure or major changes to the Tor network, while application-level defenses such as HTTPOS \cite{Luo2011} require browser modifications and traffic splitting defenses such as TrafficSliver-Net involve major Tor modifications. As a result of these advantages and RegulaTor's computational simplicity, RegulaTor can be straightforwardly implemented as a pluggable transport. 

However, a pluggable transport implementation does introduce drawbacks. While PTs protect against the ISP and attackers on the client network, they do not protect against malicious bridges. Moreover, PTs require extra steps for users to set up and use, so many users are likely to remain unprotected. While a circuit padding framework exists \cite{padding}, allowing developers to inject padding cells between the client and any node within the circuit, it does not allow for real cells to be delayed. Additionally, the circuit padding framework uses adaptive padding-style \cite{Juarez2015, torspec} state machines to describe defenses. As a result, it is unable to support RegulaTor's approach of padding based on the timing of packet surges and buffering packets if necessary. Therefore, updating the circuit padding framework to support this style of defense is a remaining hurdle for the full deployment of RegulaTor.

\begin{table*}[]
    \centering
    \begin{tabular}{p{2cm}||p{2.5cm}|p{2cm}|p{2.5cm}|p{3.5cm}|p{1.2cm}}
 \hline
 \hline

 \textbf{Category} & \textbf{Defense} & \textbf{Data Overhead} & \textbf{Latency Overhead} & \textbf{Additional Requirements} & \textbf{Defeated}\\
 \hline
 \hline

 \multirow{5}{4em}{Imitation} & Traffic Morphing \cite{Wright2009} & Moderate & None & Preliminary Trace Collection & Yes\\
 & Decoy pages\cite{Panchenko2011} & High & None & Loads another web page simultaneously & Yes\\
 & Supersequence \cite{Wang2014a} & Moderate & None & Knowledge of other traces & No\\
 & Glove \cite{Nithyanand2014} & High & High & Knowledge of other traces & No\\
 & Walkie-Talkie \cite{Wang2017} & Moderate & Moderate & Knowledge of burst information, half-duplex & Yes\\
 \hline

 \multirow{4}{4em}{Regulation} & BuFLO \cite{Dyer2012} & Very High & Very High & None & No\\
 & CS-BuFLO \cite{Cai2014a} & Very High & Very High & None & No\\
 & Tamaraw \cite{Cai2014} & Very High & Very High & None & No\\
 & WTF-PAD \cite{Juarez2015} & Moderate & None & None & Yes\\
 
 \hline
 
 \multirow{2}{4em}{Alteration} & HTTPOS \cite{Luo2011} & Low & None & Browser modifications & Yes\\
 & FRONT \cite{Gong2020} & Moderate & None & None & Partially\\
 \hline
 \multirow{3}{4em}{Traffic Splitting} & Multihoming \cite{multihoming} & None & None & multipath-compatible bridge & No\\
 & TrafficSliver-App \cite{trafficsliver} & None & Low & Uses local proxy & No\\
 & TrafficSliver-Net \cite{trafficsliver} & None & Low & Tor modifications & No\\

 \hline
 Other & Glue \cite{Gong2020} & Moderate & None & User behavior assumptions, trace database & No\\
 \hline

\end{tabular}
    \vspace{5pt}
    \captionsetup{justification=centering,margin=2cm}
    \caption{Summary of WF Defenses}
    \label{tab:defenses_summary}
\end{table*}
\section{Related Work}
\subsection{Website Fingerprinting Attacks}

While we use Tik-Tok and Deep Fingerprinting to evaluate WF defenses in this paper, they are not the only deep learning-based techniques that demonstrate high effectiveness. For example, Automated Website Fingerprinting by Rimmer et al. \cite{Rimmer2018} collected a large data set and trained three models: stacked denoising autoencoders, convolutional neural networks, and long short-term memory networks. In addition, Oh et al. 
presented p-FP \cite{Oh2019}, which demonstrates that deep neural networks can be used to generate feature vectors to improve previous WF attacks in a variety of settings, and Bhat et al. released Var-CNN \cite{Bhat2019}, which achieves high accuracy while using a reduced data set.

Other works have primarily aimed to improve existing WF attacks. These include Triplet Fingerprinting \cite{triplet}, which uses N-shot learning to allow attackers to train effective WF models on data sets of limited size, and GANDaLF \cite{gandalf}, which uses generative adversarial networks to generate "fake" data. Both of these techniques demonstrate that WF attacks can be carried out without the large data sets used in earlier deep learning-based attacks. Lastly, Pulls et al. \cite{oracle} have shown that the Website Oracle security notion can be combined with website fingerprinting attacks to greatly reduce the false positive rate for most websites visited over Tor. 

\subsection{Website Fingerprinting Defenses}

To defend against WF attacks, WF defenses alter traffic to reduce the amount of information leaked about the associated web page. Defenses typically do this by adding `dummy' packets and inserting delays into the traffic. For clarity, we categorize previously published defenses into four broad categories, which we call `imitation,' `regulation,' `alteration,' and `traffic splitting.' A summary of published defenses organized by category is given in table \ref{tab:defenses_summary}.

\textit{Imitation} defenses attempt to make packet sequences appear similar or identical to sequences of packets associated with other potential destinations, preventing an attacker from distinguishing between them. While these defenses are often effective with moderate bandwidth overhead, they require information about other packet sequences to imitate. As a result, they require the implementation of additional infrastructure or the storage of data about a user's previously visited web pages, which is potentially difficult. Still, several instances of this defense type have been presented, starting with Traffic Morphing by Wright et al. \cite{Wright2009}. This defense used convex optimization techniques to add packets and alter packet lengths to match other packet sequences, which had been previously stored by the client. However, Traffic Morphing is ineffective on Tor, which pads each `cell' to a constant size. Later, Panchenko et al. \cite{Panchenko2011} presented a method to `camouflage' a trace by loading another `decoy' simultaneously with the original one. While this approach appeared relatively effective at first, it has since been defeated by WF attacks. 

Another defense, Supersequence, \cite{Wang2014a} calculates `anonymity sets,' which are groups of similar packet sequences that are created to minimize the overhead needed to pad them such that an attacker cannot distinguish between them. Then, the shortest common supersequence is calculated to carry out the padding efficiently. The Glove defense \cite{Nithyanand2014} uses a similar strategy of clustering web pages using k-medoids and dynamic time warping for distance matrix calculation. Then, it creates `super-traces' to cover the packet sequences in their respective clusters. While these defenses are theoretically effective, they require knowledge of the full trace being defended and information about packet sequences associated with other web pages. 

The most recent imitation defense, Walkie-Talkie, \cite{Wang2017}, provides a high degree of security with moderate latency and bandwidth penalties. Its approach is to operate Tor Browser in half-duplex mode, meaning that the browser is either loading a page or making requests, but does not interleave these states. While visiting a web page, the defense uses `burst molding' to alter the packet bursts to match a packet sequence from another web page. Walkie-Talkie also partially side-steps concerns about knowledge and storage of other packet sequences by storing only burst-level features. Walkie-Talkie is effective against all defenses except for Tik-Tok \cite{Rahman2019}, which uses timing information to distinguish defended packet sequences.

\textit{Regulation} defenses attempt to make WF attacks difficult or impossible by regulating the packet sending. The original regulation defense, BuFLO \cite{Dyer2012}, was primarily created to demonstrate that such a WF defense was possible, albeit inefficient. BuFLO operates by sending packets at fixed intervals for a set length of time. If no packet is available at the set time, then a dummy one is sent instead. Later, CS-BuFLO \cite{Cai2014a} was presented as an improved version of BuFLO that adapts its transmission rate to reduce overhead and congestion. Another BuFLO variant, Tamaraw \cite{Cai2014} reduced overhead while functioning as a ``theoretically provable'' BuFLO. While these defenses are resistant to WF attacks and require no additional infrastructure or information about other packet sequences, their latency and bandwidth penalties are too high for practical use. 

Still, WTF-PAD \cite{Juarez2015} manages to efficiently regularize some aspects of the packet sequence by using an approach based on adaptive padding \cite{Shmatikov2006}, which was developed to prevent end-to-end traffic analysis. It does this by filling long gaps in packet sequences whenever the gap between packets is larger than the time length sampled from a distribution of typical inter-arrival times. By reducing the amount of information leaked by each trace while imposing only a moderate bandwidth penalty, WTF-PAD is a strong practical defense; accordingly, WTF-PAD is re-evaluated for comparison in this paper. 

\textit{Alteration} defenses aim to change packet sequence features to confuse attackers and make it likely that defended sequences are misclassified. One early example tested in Tor was HTTP pipelining, which consists of multiple HTTP requests sent with a single TCP connection \cite{experimental}. The HTTPOS defense \cite{Luo2011}, designed to prevent information leaks from encrypted flows, uses HTTP pipelining along with a variety of other browser changes, such as dummy requests, altered TCP window sizes, multiple TCP connections, and HTTP Range requests. Tor also tested a browser that requested embedded objects randomly. Still, these application-level defenses were defeated by Cai et al. \cite{Cai2012}. 

Another alteration defense is the recently-published FRONT \cite{Gong2020}, which focuses on adding dummy packets to the front of packet sequences, where most of the information about the associated web page is leaked. It also heavily emphasizes randomized padding by randomly choosing both the distribution and the volume of added packets. As a result, the packet sequences associated with any given web page often look completely different from one another in terms of both total packet volume and location of packet bursts, making it difficult for the WF attacker to accurately categorize packet sequences. FRONT requires no additional latency, a moderate bandwidth penalty, and no data collection or infrastructure. Still, it manages to be one of the most effective of the practical defenses, making it a primary target of comparison in this paper. 

\textit{Traffic Splitting} defenses defend against WF attacks by splitting traffic between multiple guard nodes so that any individual sub-trace reveals little information about the target web page. The TrafficSliver \cite{trafficsliver} and Multihoming \cite{multihoming} defenses both use traffic splitting, though their implementations are fairly different. In Multihoming, clients connect through two different access points (e.g. home WiFi and public WiFi) and then merge traffic at a multipath-compatible Tor bridge. However, TrafficSliver presents two unique approaches known as TrafficSliver-App and TrafficSliver-Net. TrafficSliver-Net alters the Tor network to split TCP traffic over multiple entry nodes and merges traffic at the middle node, while TrafficSliver-App creates multiple Tor circuits and proxies HTTP requests over the circuits. 

Both TrafficSliver and Multihoming can be implemented with minimal bandwidth and appear to be effective WF defenses. However, they do not guard against all attacker types: TrafficSliver only prevents WF attacks that take place at the guard node, and Multihoming does not protect against local attackers who can see outgoing traffic. As a result, more investigation into these limitations is needed to further develop traffic splitting attacks.

\section{Conclusion and Future Work}

In this paper, we presented a novel and lightweight WF defense, RegulaTor, along with the insights that allow for it to operate effectively without large bandwidth or latency penalties. By shaping the download packet sequences to contain large packet surges that quickly decay, the packet sequences become difficult to distinguish. Then, because the upload packet sequences closely mimic the download packet sequences, RegulaTor can send upload packets as a function of the download packet timing without incurring high latency or bandwidth overhead. Thus, the download packet sequence is regularized, and the upload packet sequence leaks no further information about the associated web page. 

Then, we re-evaluated comparable lightweight WF defenses against state-of-the-art WF attacks to demonstrate that RegulaTor provides substantially improved defense with similar or lessened overhead in both the open-world and closed-world settings. To be specific, it reduces accuracy in the closed-world setting to just 25.4\%, compared to 66.0\% for FRONT-2500, the best published deployable defense. Furthermore, it achieves this performance with only 79.7\% bandwidth overhead and a small latency overhead, while FRONT-2500 requires 119\% bandwidth overhead. Most importantly, however, it prevents state-of-the-art WF attacks from detecting visits to monitored web pages with useful accuracy, demonstrated by the precision-tuned Tik-Tok attack receiving an $F_1$-score of only .135 (compared to .625 for FRONT-2500). Thus, RegulaTor represents a step forward in terms of deployable WF defenses. 

While RegulaTor prevents an attacker from linking a defended packet sequence with an associated web page, we did not test it in terms of preventing an attack from identifying a web \textit{site} based on a series of page loads. This task may be easier for an attacker able to use information about multiple web pages on the same website as well as information about how users typically interact with these websites. Thus, defending against this type of WF attack is left for future work. Other potentially interesting settings for further analysis include testing RegulaTor with only packet sequences collected using the fastest or slowest circuits, varying the size of the unmonitored or unmonitored sets in the open-world attack, and varying the size of the data set in the closed-world setting.

\section*{Acknowledgments}
We’d  like  to  thank  Sirinam  et  al.  for  providing  their data sets. This work was funded by a 3M fellowship and NSF grant 1815757. 

\bibliographystyle{plain}
\bibliography{main}

\end{document}